
\input epsf

\ifx\epsffile\undefined\message{(FIGURES WILL BE IGNORED)}
\def\insertplot#1#2{}
\def\insertfig#1#2{}
\else\message{(FIGURES WILL BE INCLUDED)}
\def\insertplot#1#2{
\midinsert\centerline{{#1}}\vskip0.2truein
\centerline{{\epsfxsize=\hsize
\epsffile{#2}}}\vskip0.5truecm\endinsert}
\def\insertfig#1#2{
\midinsert\centerline{{\epsfxsize=\hsize\epsffile{#2}}}\vskip0.2truein
{{#1}}\vskip0.5truecm\endinsert}
\fi

\input harvmac
%
%
%
%
\ifx\answ\bigans
\else
\output={
  \almostshipout{\leftline{\vbox{\pagebody\makefootline}}}\advancepageno
}
\fi
%
%
%

%
%

%
%
\def\UCSD#1#2{\noindent#1\hfill #2%
\bigskip\supereject\global\hsize=\hsbody%
\footline={\hss\tenrm\folio\hss}}
%
%
\def\abstract#1{\centerline{\bf Abstract}\nobreak\medskip\nobreak\par #1}
%
%
%
%
\edef\tfontsize{ scaled\magstep3}
 \tfontsize  \tfontsize
 \tfontsize \font\titlei=cmmi10 \tfontsize
\font\titleis=cmmi7 \tfontsize \font\titleiss=cmmi5 \tfontsize
\font\titlesy=cmsy10 \tfontsize \font\titlesys=cmsy7 \tfontsize
\font\titlesyss=cmsy5 \tfontsize  \tfontsize
\skewchar\titlei='177 \skewchar\titleis='177 \skewchar\titleiss='177
\skewchar\titlesy='60 \skewchar\titlesys='60 \skewchar\titlesyss='60
%
%
%
%
%
\def\inv{^{\raise.15ex\hbox{${\scriptscriptstyle -}$}\kern-.05em 1}}
\def\lbar{{\lower.35ex\hbox{$\mathchar'26$}\mkern-10mu\lambda}} 

%
%
%
%
\def\dsl{\,\raise.15ex\hbox{/}\mkern-13.5mu D} 
\def\delsl{\raise.15ex\hbox{/}\kern-.57em\partial}
\def\Ksl{\hbox{/\kern-.6000em\rm K}}
\def\Asl{\hbox{/\kern-.6500em \rm A}}
\def\Dsl{\hbox{/\kern-.6000em\rm D}} 
\def\Qsl{\hbox{/\kern-.6000em\rm Q}}
\def\gradsl{\hbox{/\kern-.6500em$\nabla$}}
%
%
\def\lspace{\ifx\answ\bigans{}\else\qquad\fi}
\def\lbspace{\ifx\answ\bigans{}\else\hskip-.2in\fi} 
%
%
\def\boxeqn#1{\vcenter{\vbox{\hrule\hbox{\vrule\kern3pt\vbox{\kern3pt
        \hbox{${\displaystyle #1}$}\kern3pt}\kern3pt\vrule}\hrule}}}
%
%
\def\mbox#1#2{\vcenter{\hrule \hbox{\vrule height#2in
\kern#1in \vrule} \hrule}}
%
%
%
%

%
%
%
%
%

%

%
%

\def\darr#1{\raise1.5ex\hbox{$\leftrightarrow$}\mkern-16.5mu #1}

%
%
\def\frac#1#2{{\textstyle{#1\over #2}}} 
%
%
%
%

%
%
%
%

%
%
\def\ltap{\ \raise.3ex\hbox{$<$\kern-.75em\lower1ex\hbox{$\sim$}}\ }
\def\gtap{\ \raise.3ex\hbox{$>$\kern-.75em\lower1ex\hbox{$\sim$}}\ }
\def\gl{\ \raise.5ex\hbox{$>$}\kern-.8em\lower.5ex\hbox{$<$}\ }
\def\roughly#1{\raise.3ex\hbox{$#1$\kern-.75em\lower1ex\hbox{$\sim$}}}
%
%

%

%
\def\np#1#2#3{{Nucl. Phys. } B{#1} (#2) #3}
\def\pl#1#2#3{{Phys. Lett. } {#1}B (#2) #3}

\def\physrev#1#2#3{{Phys. Rev. } {#1} (#2) #3}

\relax

\def\lta{\ \hbox{\raise.55ex\hbox{$<$}} \!\!\!\!\!
\hbox{\raise-.5ex\hbox{$\sim$}}\ }
\def\gta{\ \hbox{\raise.55ex\hbox{$>$}} \!\!\!\!\!
\hbox{\raise-.5ex\hbox{$\sim$}}\ }

\def\qsl{\hbox{/\kern-.5600em {$q$}}}
\def\ksl{\hbox{/\kern-.5600em {$k$}}}

\def\({\left(}
\def\){\right)}

\def\OMIT#1{}
\def\frac#1#2{{#1\over#2}}

\def\lamstar{\Lambda^*}
\def\glam{g_{\Lambda^*}}
\def\glamnk{g_{\Lambda^*}(NK)}
 \def\glamsigpi{g_{\Lambda^*}(\Sigma\pi)}

\hbadness=10000

\noblackbox
\vskip 1.in
\centerline{{\titlefont{$\Lambda (1405)$ Contribution to Kaon-Nucleon}}}
\medskip
\centerline{{\titlefont{ Scattering Lengths In Chiral Perturbation Theory}
\footnote{*}{{\tenrm Work
supported in part by the Department of Energy under contract
DE--FG02--91ER40682 (CMU) and by the NSF under grant PHY89-04035.}}}}
\vskip .5in
\centerline{Martin J. Savage}
\medskip
\centerline{\it Department of Physics, Carnegie Mellon University,
Pittsburgh PA 15213}
\medskip
\centerline{\it and }
\medskip
\centerline{\it Institute for Theoretical Physics, University of California,
Santa Barbara,
CA 93106-4030.}

\vskip .2in

\abstract{
We examine the role of the $\Lambda(1405)$ in kaon-nucleon scattering lengths
using chiral
perturbation theory.
The leading nonanalytic SU(3) corrections reduce the coupling of the $\Lambda
(1405)$ to
$KN$  compared to $\Sigma\pi$.
S-wave $K^-p$ scattering is the only channel significantly affected by the $
\Lambda
(1405)$ pole which substantially cancels against the leading term fixed by the
vector
current.    This cancellation and the close proximity of the $\Lambda (1405)$
pole makes
this SU(3) correction to $a(K^-p)$  large  and  at lowest order leaves  the
sign of
$a(K^-p)$ undetermined.
We extract two linear combinations of constants appearing at higher order in
the chiral
expansion from measurements of $KN$ scattering lengths.
These constants are important for the offshell behaviour of $KN$ scattering
amplitudes
that play a central role in kaon condensation.
The recent claims about the calculability of these constants are discussed.
 }

\vfill
\UCSD{\vbox{
\hbox{CMU-HEP 94-11}
\hbox{NSF-ITP-94-33}
\hbox{DOE-ER/40682-65}}
}{April 1994}
\eject

The effects of the $\Lambda(1405)$ (denoted below by $\lamstar$) with
$J^\pi = 1/2^-$ on kaon-nucleon scattering lengths have been discussed recently
\ref\weise{W. Weise, Nucl. Phys. A553 (1993) 59.}\nref\ljmr{C. Lee, H. Jung, D.
Min and M. Rho, SNUTP-93-81, (1993).}-\ref\lbr{C. Lee, G.E. Brown and M. Rho,
SNUTP-94-28, (1994).}\   in relation to the possibility of kaon condensation in
dense nuclear
matter
\ref\davann{D.B. Kaplan and A.E. Nelson, \pl{175}{1986}{57}.}\nref\davannb{D.B.
Kaplan and A.E. Nelson, \pl{192}{1987}{193}.}\nref\bkrt{G.E Brown, K. Kubodera,
M.
Rho and V. Thorsson, \pl{291}{1992}{355}.}\nref\bkr{G.E. Brown, K. Kubodera and
M.
Rho, \pl{192}{1987}{273}.}\nref\pw{H.D. Politzer and M.B. Wise,
\pl{273}{1991}{156}.}\nref\mpw{D. Montano, H.D. Politzer and M.B. Wise,
\np{375}{1992}{507}.}\nref\blrt{G.E Brown, C.H. Lee, M. Rho and V. Thorsson,
SUNY-NTG-93-7 (1993).}\nref\dee{J. Delorme, M. Ericson and T.E.O. Ericson,
\pl{291}{1992}{379}.}\nref\ynmk{H. Yabu, S. Nakamura, F. Myhrer and K.
Kubodera,
\pl{315}{1993}{17}.}\nref\tpl{V. Thorsson, M. Prakash and J.M. Lattimer
NORDITA-93/29 N (1993).}-\ref\lsw{M. Lutz, A. Steiner and W. Weise, TPR-93-19
(1993).}. The question of whether a kaon condensate forms requires knowledge of
the
scattering amplitudes of offshell kaons on nucleons at finite density. It is
desirable to make a
model independent investigation of this phenomenon using only the symmetries of
QCD,
exploited by the chiral lagrangian.  By construction, there will be unknown
quantities that
must be determined by comparison with experimental data.
We will investigate how the $KN$ scattering lengths as presently determined
constrain
the offshell behaviour of these scattering amplitudes. The
$\lamstar$ spin and parity are such that
$K^-p$ scattering through this resonance occurs in an S-wave and its close
proximity to the
$K^-p$ threshold ($\sim 30 {\rm MeV}$) requires that it be included in any
consistent
analysis of kaon-nucleon scattering (for the role of the $\lamstar$ on kaonic
atoms see  e.g.
\ref\bulo{H. Burkhardt and J.Lowe, \physrev{C44}{1991}{607}.}
and references therein.).
Furthermore, its contribution to $K^-p$ scattering substantially cancels
against the
leading  term arising from the vector current.
The pole enhancement of the $\lamstar$  contribution and the large cancellation
between terms suggests that small changes in the $\lamstar$ coupling to $KN$
may be
important.
In this letter we compute the leading SU(3) violation in the $\lamstar$
coupling to octet
baryons in chiral perturbation theory and their effects on kaon-nucleon
scattering lengths.
Also, at lowest order we extract a linear combination of unknown constants in
the offshell scattering amplitudes from the onshell $K^+N$ scattering lengths
and
discuss the recent claims  \ljmr\lbr\blrt\ about calculability of these higher
order terms in the
chiral expansion.

The dynamics of the pseudo-Goldstone bosons associated with the spontaneous
breaking of
$SU(3)_L\otimes SU(3)_R$ chiral symmetry to $SU(3)_V$  are
described by the lagrange density
\eqn\meson{ {\cal L}_\pi = {f^2\over 8}Tr[
\partial_\mu\Sigma^\dagger\partial^\mu\Sigma ]
 + \mu Tr[ m_q\Sigma + {\rm h.c.} ] + .....\ \ \ ,}
where $\Sigma = \exp\left( 2iM/f\right)$  is the exponential of the meson field
where
\eqn\mesmat{ M = \left(\matrix{ \eta/\sqrt{6} + \pi^0/\sqrt{2}&\pi^+& K^+\cr
\pi^-&\eta/\sqrt{6} - \pi^0/\sqrt{2}&\overline{K}^0\cr
K^-&K^0&-2\eta/\sqrt{6}}\right)\ \ \ ,}
$f$ is the meson decay constant ($f_\pi  = 132$MeV) and $m_q$ is the light
quark mass
matrix. The dots denote terms higher order in the chiral expansion containing
more
derivatives or more insertions of the light quark mass matrix.
The lowest order chiral lagrangian describing
low momentum interactions of the pseudo-Goldstone bosons with the lowest lying
baryons of velocity $v$  is (for a review see
\ref\jmhung{E. Jenkins and A.V. Manohar, Proceedings of the workshop on {\it
Effective
field Theories of the Standard Model}, edited by U. Meissner (World
Scientific), Singapore
(1992).})
\eqn\octet{\eqalign{  {\cal L}_B =  &\  iTr[ \overline{B}_v v\cdot {\cal D} B_v
]
-i \overline{T}_v^\mu v\cdot {\cal D} T_{v\mu}
+i\overline{\Lambda^*_v} v\cdot \partial \Lambda^*_v
+\Delta_T \overline{T}_v^\mu T_{v\mu}
-\Delta_{\Lambda^*}\overline{\Lambda^*_v}\Lambda^*_v           \cr
&\ + 2DTr[\overline{B}_v S^\mu_v \{ A_\mu , B\}]
+ 2FTr[\overline{B}_v S^\mu_v [ A_\mu , B]]
+{\cal C}\left( \overline{T}_v^\mu
A_\mu B_v + {\rm h.c.} \right)  \cr
&\ +2{\cal H}\overline{T}_v^\mu S_\nu A^\nu T_{v\mu}
+ g_{\Lambda^*}\left( \overline{\Lambda^*_v} Tr[ v\cdot A B_v ] + {\rm
h.c.}\right)
} \ \ \ ,}
where $S^\mu$ is the spin operator,
$A_\mu = {i\over 2}\left( \xi\partial_\mu\xi^\dagger -
\xi^\dagger\partial_\mu\xi \right)$
is the meson axial current  ($\xi = \exp\left(iM/f\right)$)
and
${\cal D}B_v = \partial B_v  + [V ,  B_v  ]$ where
$V_\mu = {1\over 2}\left( \xi\partial_\mu\xi^\dagger +
\xi^\dagger\partial_\mu\xi \right)$
is the meson vector current.
The octet baryons are denoted by the field
\eqn\ocbary{ B_v  =  \left(\matrix{ \Lambda_v/\sqrt{6} +
\Sigma_v^0/\sqrt{2}&\Sigma_v^+& p_v\cr
\Sigma_v^-&\Lambda_v/\sqrt{6} - \Sigma_v^0/\sqrt{2}&n_v\cr
\Xi_v^-&\Xi_v^0&-2\Lambda_v/\sqrt{6}}\right)\ \
\ .}
and the decuplet of baryon resonances is denoted by
the field $T_v^\mu$.
The axial couplings $F, D, {\cal C}$  and ${\cal H}$ are determined
from semileptonic decays of the octet baryons
\ref\jma{E. Jenkins and A.V. Manohar, \pl{255}{1991}{558}.}
\ref\jmb{E. Jenkins and A.V. Manohar, \pl{259}{1991}{353}.}
and from the strong decays of the decuplet baryons
\ref\bss{M.N. Butler, M.J. Savage and R.P Springer, \np{399}{1993}{69}.}.
The $\Lambda(1405)$ $J^\pi={1\over 2}^-$ resonance of four-velocity $v$ is
denoted by
$\lamstar_v$ (its inclusion has been discussed by \ljmr \foot{Our definition of
$\glam$ in
\octet\ is a factor of\  $\sqrt{2}$\  larger than that defined in \ljmr\ .}).
It has S-wave coupling to a pseudo-Goldstone boson and octet baryon (denoted by
$\glam$) whereas it has only D-wave couplings to a pseudo-Goldstone boson  and
decuplet
baryon, which we will neglect in the work.
The mass difference between the octet baryon and the decuplet baryons is
denoted by
$\Delta_T$ and the mass difference between the octet baryons and the $\lamstar$
is
denoted by $\Delta_{\Lambda^*}$.

The width of the $\lamstar$ is dominated by the strong
decay $\lamstar\rightarrow\Sigma\pi$
determined by $\glam$ at lowest order in the chiral expansion.
In the limit of exact flavour SU(3) the couplings $\glam$, $\glamsigpi$ ( the
coupling to
$\Sigma\pi$ ) and $\glamnk$  ( the coupling to $NK$ ) are all equal.  However,
SU(3)
breaking arising from the difference between the mass of the strange quark and
the mass of
the up and down quarks gives rise to a difference between these couplings.  As
$\glamnk$ is
the coupling constant relevent for $KN$ scattering it is important to estimate
the size of this
SU(3) breaking. The leading nonanalytic SU(3) violations to the $\lamstar$
couplings arise
from the graphs shown in
\fig\lamcoup{The Feynman diagrams giving rise to the leading SU(3) violation in
the
coupling of the $\Lambda(1405)$ to octet baryons and pseudo-Goldstone bosons.
The dashed lines denote pseudo-Goldstone bosons and the solid lines denote
baryons.
Graphs (a) and (b) are wavefunction renormalisations for the mesons and baryons
respectively and graphs (c) and (d) are vertex corrections.}.
As the $\lamstar$ is not an asymptotic state we do not include its wavefunction
renormalisation in the computation (if we were to treat it as an asymptotic
state then
as it is an SU(3) singlet its wavefunction renormalisation does not contribute
to SU(3)
violating observables at lowest order.) .  The one-loop coupling constant
$g_{\Lambda^*}(BM)$ (coupling of the
$\lamstar$ to an octet baryon $B$ and pseudo-Goldstone boson $M$) is
\eqn\coup{g_{\Lambda^*} (BM) = g_{\Lambda^*}\left[ \vphantom{3\over 2}
 1 +  O_M + O_B + O_V(BM) +  O_A(BM) \right] \ \ \ \ .}
$O_M$ are the contribution from the wavefunction
renormalisation of the meson fields \lamcoup (a) and are
\eqn\mespsi{\eqalign{ O_\pi = &\ {2\over
96\pi^2f^2}M_K^2\log\left(M_K^2/\Lambda_\chi^2\right)
 \cr O_K = &\ {5\over
96\pi^2f^2}M_K^2\log\left(M_K^2/\Lambda_\chi^2\right)  }\ \ \ ,}
while the
contribution from the wavefunction renormalisation of the octet baryons
\lamcoup (b) can be
found in
\jmhung, and are
\eqn\waveoct{\eqalign{ O_\Sigma = &\ -{1\over32\pi^2f^2}
M_K^2\log\left(M_K^2/\Lambda_\chi^2\right)
\left[ {26\over 3}D^2 + 6F^2 + {14\over3}{\cal C}^2 \right]\cr  O_N = &\
-{1\over32\pi^2f^2} M_K^2\log\left(M_K^2/\Lambda_\chi^2\right)
\left[ {17\over 3}D^2 + 15F^2 - 10FD + {\cal C}^2 \right]}\ \ \ \ .}
We have only retained the nonanalytic contributions from the $K$ and $\eta$
loops using $M_\eta^2 = {4\over 3}M_K^2$ and $\Lambda_\chi$ is the chiral
symmetry breaking scale.
For the purposes of this calculation we have neglected the mass splitting
between the octet
and decuplet baryons $\Delta_T=0$.
The contributions from the graphs involving the
two-meson vector coupling arising from the octet baryon kinetic energy term
\lamcoup (c) are
\eqn\vect{\eqalign{ O_V(\Sigma\pi) = &\  {1\over
32\pi^2f^2}\left(\vphantom{3\over 2}
 M_K^2 F_\beta (\Xi K)
+ M_K^2 F_\beta (N K)  -4 M_\pi^2 F_\beta (\Sigma\pi) \right)
\cr O_V(NK) = &\  {1\over 32\pi^2f^2}\left( 3M_K^2 F_\beta (\Xi K)
+{3\over 2}M_\eta^2 F_\beta (\Lambda\eta)   +{3\over 2} M_\pi^2
F_\beta (\Sigma\pi) \right) }\ \ \ \ ,} where the function
$F_\beta(BM)$ is shorthand for
$F_\beta ( (M_{\Lambda^*}-M_B)/M_M , M_M/\Lambda_\chi )$ and
\eqn\fbetdef{ F_\beta(y,z) = (1-y^2)\log z^2  +
2y\sqrt{y^2-1}\log\left({-y+\sqrt{y^2-1+i\epsilon}\over
-y-\sqrt{y^2-1+i\epsilon} }\right) \
\ \ \ ,}  where we have only retained the leading nonanalytic terms.
The last contributions arise from the axial coupling to three mesons \lamcoup
(d) and are
\eqn\axial{\eqalign{ O_A(\Sigma\pi) = &\ -{2\over 96\pi^2 f^2}M_K^2\log\left(
M_K/\Lambda_\chi^2\right)  \cr
O_A(NK) = &\ -{5\over 96\pi^2 f^2}M_K^2\log\left(M_K/\Lambda_\chi^2\right)  }
\ \ \ \ ,}
which is an equal and opposite contribution to that from ${\cal O}_M$ (this is
also true for
the contributions from the $\pi$'s).

The observed width of the $\Lambda(1405)$ of $\Gamma_{\Lambda^*} = 50\pm 2$
\ref\pdg{Particle Data Group, \physrev{45}{1992}{1}.}
leads to a value of $|g_{\Lambda^*} (\Sigma\pi)| = 0.58 \pm 0.01$ for
$f=f_\pi$.
Using $D=0.7\pm 0.2$, $F=0.5\pm 0.1$, ${\cal C}=-1.2\pm 0.2$ and
$\Gamma_{\Lambda^*}$ as  input parameters for \coup\ we find that  (up to an
overall
sign)
\eqn\central{\eqalign{
g_{\Lambda^*}  = &\ 0.40 \pm 0.04 \cr
g_{\Lambda^*} (\Sigma\pi) = &\  (0.58 \pm 0.01) + (0.12 \pm 0.01) i \cr
g_{\Lambda^*} (NK) = &\  (0.32 \pm 0.03) + (0.050 \pm 0.005) i }\ \ \ \ .}
The imaginary parts of the couplings arise from physical intermediate states in
the loop
graphs shown in \lamcoup\ .
The $\lamstar$ coupling to $NK$ is seen to be suppressed relative to its
coupling to
$\Sigma\pi$.
It should be remembered that only the leading
nonanalytic corrections to $\glamnk$ and $\glamsigpi$ have been included and
therefore
only part of the SU(3) violation. However, as discussed in
\jmhung\bss\ , the leading nonanalytic corrections appear to capture most of
the SU(3)
violation in axial current matrix elements between octet and/or decuplet
baryons and we
hope that this is true for these couplings also.
The above uncertainties are those associated with the input parameters only.
There are
systematic uncertainties that are not included arising from truncation of the
chiral
expansion.

At leading order in the $1/M_N$ expansion ( $M_N$ is the nucleon mass ) and up
to order
$Q^2$ where $Q=\partial$, $m_s^{1/2}$  the kaon-nucleon S-wave scattering
lengths  are
\eqn\lengths{\eqalign{  a(K^\pm p) = &\  {1\over 4\pi}{M_N\over M_N+M_K}\left(
\mp {2M_K\over f^2} - {g_{\Lambda^*}^2M_K^2\over f^2}{1\over
\mp M_K-\Delta_{\Lambda^*} +i{\Gamma\over 2}} + C_p\right)  \cr a(K^\pm n) = &\
{1\over 4\pi}{M_N\over M_N+M_K}\left(  \mp {M_K\over f^2}+ C_n\right)  } \ \ \
,}
where $C_{n,p}$ are unknown constants arising  at order $Q^2$ which we will
discuss
subsequently (at order $Q^3$ and higher the constant $C_p$ for $K^+$ scattering
will be
different  from that for $K^-$ scattering but at order $Q^2$ they are equal).
The factor of $M_N/(M_N+M_K)$ is kinematic in origin and is retained
explicitly.
It comes as no surprise that a new scale, $(M_K-\Delta_\lamstar)\sim 30 {\rm
MeV}$, has
appeared in the $K^-p$ channel from the $\lamstar$ resonance, the difference
between the
position of the pole and the $K^-p$ threshold. Despite the appearance of
$M_K^2$ in the
contribution from the $\lamstar$ suggesting an order $Q^2$ term, the
effect of the pole  makes it the same size as the leading $M_K$ term in the
$K^-p$
channel.  Also the finite width of the
$\lamstar$ is not higher order for this particular process as its effects are
enhanced
by this new small scale.

The experimentally determined scattering length for $K^+p$ is
$a(K^+p) = -0.31\pm 0.02 {\rm fm}$ (for a review of the available data see
\ref\bs{T. Barnes  and E.S. Swanson, \physrev{C49}{1994}{1166}.}
\ref\harw{J.S. Hislop, R.A. Arndt, L.D. Roper and R.L Workman,
\physrev{46}{1992}{961}.}.
The uncertainty in $a(K^+p)$ is an estimate based on fig.2 in
\bs\ ). If we set $C_p=0$ in expression \lengths\ then we would find
\eqn\cpzero{a(K^+p; C_p=0, g_{\Lambda^*}) = -0.56\pm 0.01 \ {\rm fm}\ \
\ \ ,} which is  insensitive to $g_{\Lambda^*}$ and its associated
uncertainties as it is far from the $\lamstar$ pole ($f=f_\pi$).
Neglecting the dependence upon  $g_{\Lambda^*}$ we find that
\eqn\cptree{{1\over 4\pi}{M_N\over M_N+M_K}C_p =  0.25\pm 0.02\ {\rm fm}\ \
\ .}
Though not surprising, it is somewhat disturbing that this higher order term
contributes
$\sim 50\%$ of the leading amplitude (it is expected to be
smaller than the leading term by only
$M_K/\Lambda_\chi\sim 1/2$ from dimensional analysis) .  It
is interesting to ask about predictions we can now make about the
$K^-p$  scattering length $a(K^-p)$.
Using  \lengths\  and the central values for $C_p$ \cptree\ and
$g_{\Lambda^*}(BM)$
\central\ we find that
\eqn\kmp{\eqalign{
a(K^-p\ ;\ g_{\Lambda^*} (\Sigma\pi))  = &\  -0.31+0.43 i \ {\rm fm}\cr
a(K^-p\ ;\ g_{\Lambda^*}(NK))  = &\ 0.46+0.23i \ {\rm fm}}
\ \ \ \ ,}
which is very sensitive to the choice of $g_{\Lambda^*}$ due to the close
proximity of the
$\lamstar$ pole and the large cancellation that occurs between this term and
the tree-level
contribution.
It is clear that at this order we are not in a position to make a reliable
prediction about either
the sign or magnitude of the $K^-p$ scattering length due to the expected large
size of
corrections. This is in contrast to the results of \ljmr\ where the sign of
$Re(a(K^-p))$ is found to be negative.
Thus $a(K^-p)$ does not provide a good diagnostic with which to test
convergence of the chiral expansion and also cannot be used to determine
unknown
counterterms.

The same procedure can be applied to the kaon-neutron scattering lengths in
\lengths\ .
Scattering amplitudes for $K^+n$ are found from $K^+d$ scattering after
removing the
proton contribution (see \harw\ ).   The extracted scattering lengths for such
processes have
significant uncertainties and model dependences.  For our purposes we assume
that
$a(K^+n)$ (which has both isospin 0 and 1 contributions) lies between $-0.15\
{\rm fm}$
and
$-0.30\ {\rm fm}$ (estimated from fig.2 in \bs\ ).
We fit the constant $C_n$  from this scattering length as there are no pole
graphs that can
contribute significantly to this amplitude. If we set $C_n=0$ then we find
\eqn\akpn{a(K^+n\ ;\  C_n=0) = -0.29\  {\rm fm}\ \ \ \,}
which leads to
\eqn\cntree{-0.01\ \  {\rm fm}\ \ \ltap\  {1\over 4\pi}{M_N\over
M_N+M_K}C_n\  \ltap\   +0.14  \ \ {\rm fm}\ \ \ .}
These in turn lead to predictions for the $K^-n$ scattering lengths at leading
order of
\eqn\akpn{+0.28\ \ {\rm fm}\ \ltap\ a(K^-n) \ \ltap\  +0.43 \ \ {\rm fm}\ \ \
\,}
One might be a bit nervous
about these predictions for the same reasons that the predictions for the
$a(K^-p)$ are
unstable to higher order corrections.  However, the nearest confirmed S-wave
$I=1$
resonance is the
$\Sigma(1620)$\  ($J^\pi={1\over 2}^-$) which only weakly couples to $NK$ \pdg\
{}.
Therefore we may expect that  the meson interactions calculable in
chiral perturbation theory are the leading corrections to our results.

If we are interested in possibility of  kaon condensation in dense nuclear
matter then it is
the scattering amplitudes for offshell kaons that are required.
In order to make progress in this direction we need to understand the origin of
the constants
$C_{n,p}$ appearing in \lengths .
At order $Q^2$ the lagrange density responsible for $C_{n,p}$ is given by
\eqn\next{\eqalign{
{\cal L}_{\rm mass} =  &\ a_1Tr[ \overline{B}_v \chi_+ B_v ]
+ a_2Tr[ \overline{B}_v  B_v \chi_+ ]
+a_3Tr[  \overline{B}_v  B_v ] Tr[ \chi_+ ] \cr
&\ + d_1Tr[ \overline{B}_v  A^2 B_v   ]
+ d_2Tr[ \overline{B}_v  (v\cdot A)^2 B_v   ]
+ d_3Tr[ \overline{B}_v B_v A^2  ]\cr
&\ + d_4Tr[ \overline{B}_v B_v  (v\cdot A)^2 ]
+ d_5 Tr[ \overline{B}_v B_v   ]Tr[A^2]
+ d_6Tr[ \overline{B}_v  B_v   ] Tr[(v\cdot A)^2]\cr
&\ + d_7Tr[ \overline{B}_v  A^\mu] Tr[ A_\mu B_v   ]
+ d_8Tr[ \overline{B}_v  (v\cdot A)] Tr[(v\cdot A) B_v   ]
}\ \ \ ,}
where $a_{1,2,3}$ and $d_{1,..8}$ are unknown constants that must be determined
experimentally  and
$\chi_+ = \xi m_q\xi + \xi^\dagger m_q \xi^\dagger $.  The coefficients $a_1$
and
$a_2$ can be determined from the octet baryon mass splittings
\ref\jmsig{E. Jenkins and A.V. Manohar, \pl{281}{1992}{336}.} where it is found
that
$m_s a_1 = -65 {\rm MeV}$ and $m_s a_2 = -125 {\rm MeV}$.
$a_3$ is the``sigma term'' which does not contribute to mass splittings but can
be
measured by low energy $\pi N$ scattering where it is found to be sensitive to
kaonic loop corrections
\jmsig\ref\gls{J. Gasser, H. Leutwyler and M.E. Sainio, \pl{253}{1991}{252}.}.
It is important to stress that the coefficients $d_i$ are not calculable from
chiral
symmetry arguments alone and in particular  there is no sense in which they can
be computed
as has been claimed in
\ljmr\  and discussed in \blrt\lbr .
\foot{In \ljmr\lbr\blrt\ the contribution to the coefficients $d_i$ arising at
order $1/M_N$
are computed from the``relativistic'' lagrangian for nucleon interactions.
It is important to realise that there are operators, e.g.
$Tr[D_\mu\overline{B}_b A^\mu A^\nu D_\nu B_v]$ , that should be
included in addition to the contributions arising at order $1/M_N$.
Such operators have unknown coefficients and give a contribution of order
$\left( M_N/\Lambda_\chi\right)^2$ for nucleons near their mass shell,
in fact operators with arbitrary powers of
$D^\mu$ will give unsuppressed contributions.}
The constants $C_{p,n}$ determined by \cntree\  and \cptree\ constrains two
linear
combination of the $a_i$ and $d_i$ given by
\eqn\constrain{\eqalign{
C_p = &\ -{2m_s\over f^2}\left( a_1+a_2+2a_3\right)
+{M_K^2\over f^2}\left( d_1+d_2+d_3+d_4+2d_5+2d_6+d_7+d_8\right)\cr
C_n = &\ -{2m_s\over f^2}\left( a_2+2a_3\right) +{M_K^2\over f^2}\left(
d_3+d_4+2d_5+2d_6\right) }\ \ \ ,}
where $m_s$ is the mass of the strange quark and we have neglected the mass of
both the
up and down quarks. The constants $d_i$ have dimensions of inverse mass, and
naive
dimensional analysis would indicate $d_i\sim1/\Lambda_\chi$.
At the order to which we are working these are the only two constraints
that can be obtained from onshell $KN$ scattering.
The offshell $KN$ scattering amplitudes become a function of $a_3$ alone
(neglecting loop
and finite density effects) since the same linear combinations of
$d_i$ and $a_i$ appear both onshell and offshell and $a_{1,2}$ are fixed from
octet baryon
mass splittings.
Low energy pion-nucleon scattering is described by the lagrange densities given
in \octet
\next\ and a recent analysis of $\pi N$ scattering lengths can be found in
\ref\bkm{V. Bernard, N. Kaiser and U.G. Meissner,
\pl{309}{1993}{421}.} .
The constants $d_{3,4,7,8}$ and $a_2$ do not contribute to these scattering
amplitudes and
hence cannot be constrained by the experimentally determined $\pi N$ scattering
lengths.

In conclusion, we have computed the leading nonanalytic SU(3) violating
corrections to the
$\lamstar$ coupling constant.
We find that the $\lamstar$ coupling to $KN$ is suppressed relative to its
coupling to
$\Sigma\pi$. The substantial cancellation between the tree-level and
$\lamstar$ contributions to the $K^-p$ scattering amplitude and the close
proximity of the
$\lamstar$ pole to the $NK$ threshold enhances the effect of this SU(3)
correction.
We fit two  linear combinations of coefficients of higher dimension operators
to
$a(K^+p)$, $a(K^+n)$ and  find that the sign of $a(K^-p)$ cannot be determined
at this
order contrary to the result of  \ljmr\ .
We also stress that the coefficients of higher dimension operators  required to
compute offshell scattering amplitudes important for kaon condensation cannot
be
calculated from chiral symmetry arguments.
Corrections to our results arising from loop graphs and higher dimension
operators are
expected to be suppressed by only $M_K/\Lambda_\chi \log\left(
M_K^2/\Lambda_\chi^2\right)$ compared to the $C_{n,p}$ and may be
significant.

\bigskip\bigskip
\vfill\eject
\centerline{{\bf Acknowledgements}}
\bigskip
I would like to thank D. Kaplan, Ann Nelson and M. Veltman for useful
conversations.
I would also like to thank the Institute for Theoretical Physics at UC Santa
Barbara for
kind hospitality throughout this work.
This work is supported in part by the Department of
Energy under  contract DE-FG02-91ER40682 and in part by the NSF under grant
PHY89-04035.

\listrefs
\listfigs
\vfill\eject
\bye